\documentclass[jkps,preprint,fleqn,showpacs,showkeys]{revtex4}
\usepackage{graphicx}
\usepackage{amssymb}
\usepackage{amsmath}
\usepackage{epstopdf}
\usepackage{bm}
\usepackage{float}
\usepackage{subfigure}
\begin{document}
\setcounter{page}{1}
\title[]{An Analytical Study on the Synchronization of Strange Non-Chaotic Attractors}
\author{G. Sivaganesh}
\email{sivaganesh.nld@gmail.com}

\affiliation{Department of Physics, Alagappa Chettiar College of Engineering $\&$ Technology, Karaikudi, Tamilnadu-630 004, India}
\author{A. Arulgnanam}
\affiliation{Department of Physics, St.John's College, Palayamkottai, Tamilnadu-627 002, India}


\date[]{Received 16 September 2016}

\begin{abstract}
In this paper we present an analytical study on the synchronization dynamics observed in unidirectionally-coupled quasiperiodically-forced systems that exhibit {\emph{Strange Non-chaotic Attractors (SNA)}} in their dynamics. The SNA dynamics observed in the uncoupled system is studied analytically through phase portraits and poincare maps. A difference system is obtained by coupling the state equations of similar piecewise linear regions of the drive and response systems. The mechanism of synchronization of the coupled system is realized through the bifurcation of the eigenvalues in one of the piecewise linear regions of the difference system. The analytical solutions obtained for the normalized state equations in each piecewise linear region of the difference system has been used to explain the synchronization dynamics though phase portraits and timeseries analysis. The stability of the synchronized state is confirmed through the {\emph{Master Stability Function}}. An explicit analytical solution explaining the synchronization of SNAs is reported in the literature for the first time.

\end{abstract}

\pacs{05.45.Xt, 05.45.-a}

\keywords{Synchronization, Strange Non-chaotic Attractors, Unidirectional-Coupling}

\maketitle

\section{Introduction}

Strange non-chaotic attractors (SNAs) are geometrical structures existing between periodic and chaotic attractors. SNAs have been insensitive to initial conditions as evidenced by the negative values of their Lyapunov exponents. However, they have the complicated geometrical structure of fractals. SNAs are generic in quasiperiodically forced systems. After pioneering work of Grebogi \cite{Grebogi1984}, a large number of quasiperiodic systems were reported for the existence of SNA behavior. Different routes, such as the Heagy-Hammel or torus doubling, fractalization, intermittency, blow out bifurcation routes, to SNA have been identified \cite{Heagy1994,Venkatesan2000,Awadhesh2001,Manish2010}. A good number of nonlinear systems and electronic circuits exhibiting SNAs in their dynamics have been studied numerically and experimentally \cite{Srinivasan2009,Thamilmaran2006,Venkatesan1999,Zhong1996,Arulgnanam2015,Arulgnanam2015a,Sivaganesh2016a} while a few systems have been studied analytically \cite{Sivaganesh2014,Arulgnanam2015,Arulgnanam2015a}. The phenomenon of chaos synchronization finding potential applications in secure communication has been enchanting researchers after the {\emph{Master-Slave}} concept introduced by Pecora and Carroll \cite{Pecora1990,Carroll1991}. Different types of synchronization phenomenon, such as complete, phase, lag, anti-phase and generalized synchronization, have been identified in coupled chaotic systems. Complete synchronization of identical chaotic systems have been confirmed through the negative values of the {\emph{Master Stability Function (MSF)}}\cite{Pecora1998,Pecora1999}. The MSFs for a few nonlinear systems and simple electronic circuits have been studied \cite{Liang2009,Sivaganesh2016}. An explicit analytical solution explaining the complete synchronization of identical $Murali-Lakshmanan-Chua$ circuits has been presented \cite{Sivaganesh2015,Venkatesh2016}. Synchronization of coupled driven-damped SQUIDS exhibiting SNAs in their dynamics have been synchronized using {\emph{in-phase driving}} method \cite{Ramasamy1997}. However, the nonlinear systems exhibiting SNAs in their dynamics have not been studied analytically hitherto. Recently, the phenomenon of complete synchronization has been observed in coupled quasiperiodically-forced systems through unidirectional coupling of the systems\cite{Sivaganesh2016a}. In this paper, we present an explicit analytical solution to the complete synchronization phenomenon observed in the above said unidirectionally-coupled system exhibiting SNAs in their dynamics.   \\

This paper is divided into two sections. In Section \ref{sec:1} we present the normailized state equations of the circuit and summarize the analytical solutions of the individual circuit system through phase portraits and poincare sections. Explicit analytical solutions to the coupled state equations are presented in Section \ref{sec:2}. Further, the analytical solutions obtained in this section have been used to study the synchroniation dynamics through phase portraits. The stability of the synchronized state is confirmed through the MSF. 

\section{Circuit Equations and dynamics}
\label{sec:1}

\begin{figure}
\resizebox{0.6\textwidth}{!}{%
  \includegraphics{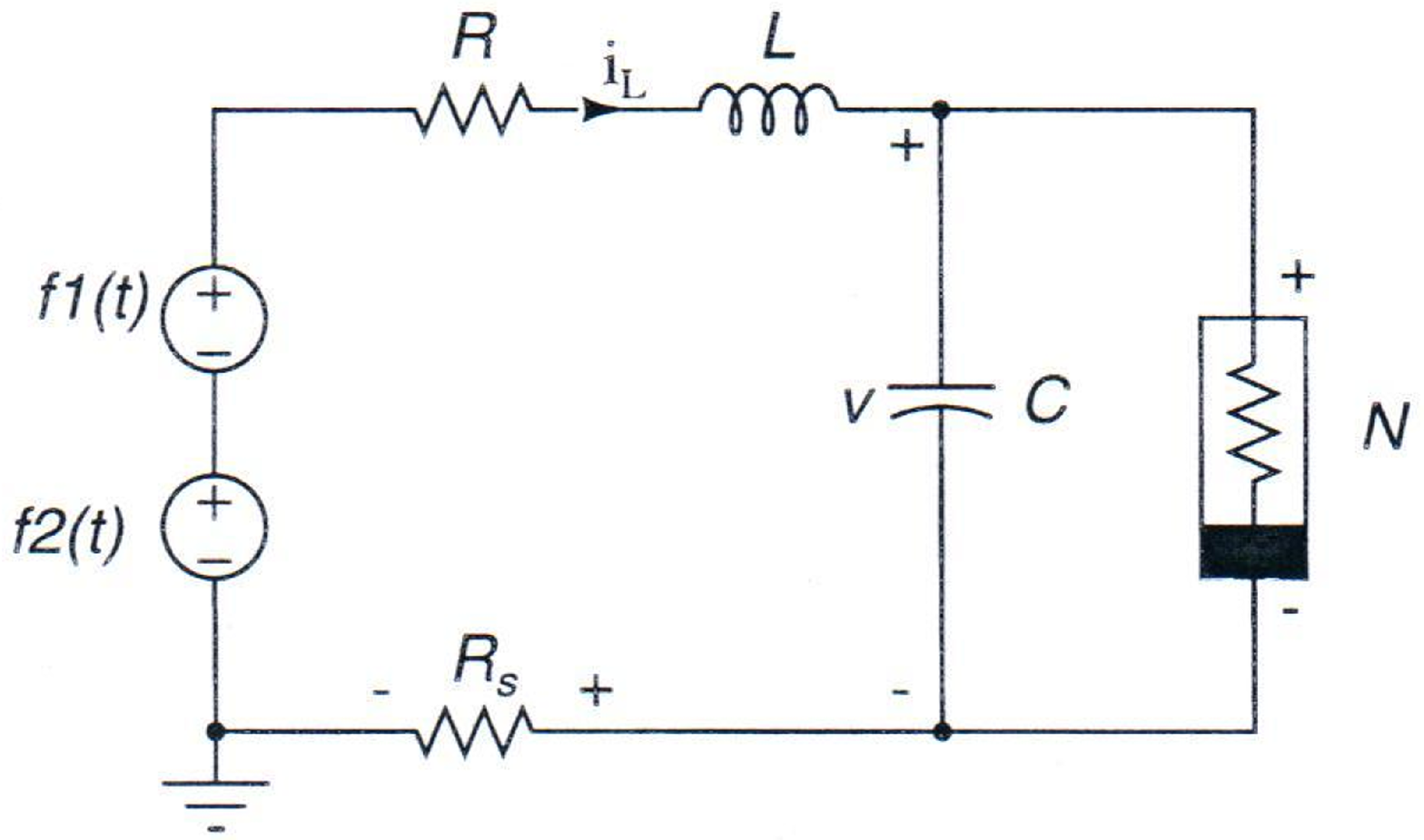}
}
\caption{Schematic circuit realization of the quasiperiodically-forced series $LCR$ circuit.}
\label{fig:1}       
\end{figure}

The circuit under present study consists of a forced, series LCR circuit with a Chua's diode connected parallel to the capacitor. The circuit is subjected to sinusoidal external forcing by two voltage sources $F_1, F_2$. A schematic diagram of the circuit is as shown in Fig. \ref{fig:1}. The state equations of the circuit is written as
\begin{subequations}
\begin{eqnarray} 
C {dv \over dt } & = & i_L - g(v), \\
L {di_L \over dt } & = & - R_t  i_L - v + F_1 sin( \omega_1 t) + F_2 sin( \omega_2 t),  
\label{eqn:1}
\end{eqnarray}
\end{subequations}
where $R_t = R + R_s$ and $g(v)$ is the piecewise linear function given by
\begin{eqnarray}
g(v) = G_bv + 0.5(G_a-G_b)[|v+B_p|-|v-B_p|],
\end{eqnarray}
where $v$ and $i_L$ being the voltage across the capacitor $C$ and the current through the inductor $L$, respectively. The value of the negative slopes of the inner, outer regions and the breakpoints in the $(v-i)$ characteristic curve of the nonlinear element are given as $G_{a} = -0.76~mS,~ G_{b} = -0.41~mS$ and $B_{p} = 1.0~V, $ respectively. With the rescaling parameters $x = v/B_p$, $y = i_L/G B_p$, $ \beta $ = $C$/$LG^2$, $ \nu $ = $GR_s$, $ f_1 $ = $F_1 \beta$/$B_p$,  $ f_2 $ = $F_2 \beta$/$B_p$, $ a $ = $G_a$/$G$,  $ b $ = $G_b$/$G$, and  $ G $ = $1$/$R$, the normalized state equations of the quasiperiodically-forced circuit can be written in the autonomous form as
\begin{subequations}
\begin{eqnarray}
\dot x  &=&  y - g(x), \\
\dot y  &=&  - \sigma y - \beta x+ f_1 sin(\theta) + f_2 sin(\phi), \\
\dot \theta  &=&  \omega_1, \\
\dot \phi &=& \omega_2,
\end{eqnarray}
\label{eqn:3}
\end{subequations}
where, $\sigma = (\beta + \nu \beta)$ and 
\begin{equation}
g(x) =
\begin{cases}
bx+(a-b) & \text{if $x\ge 1$}\\
ax & \text{if $|x|\le 1$}\\
bx-(a-b) & \text{if $x\le -1$}.
\end{cases}
\end{equation}

\begin{figure}
\resizebox{0.6\textwidth}{!}{%
  \includegraphics{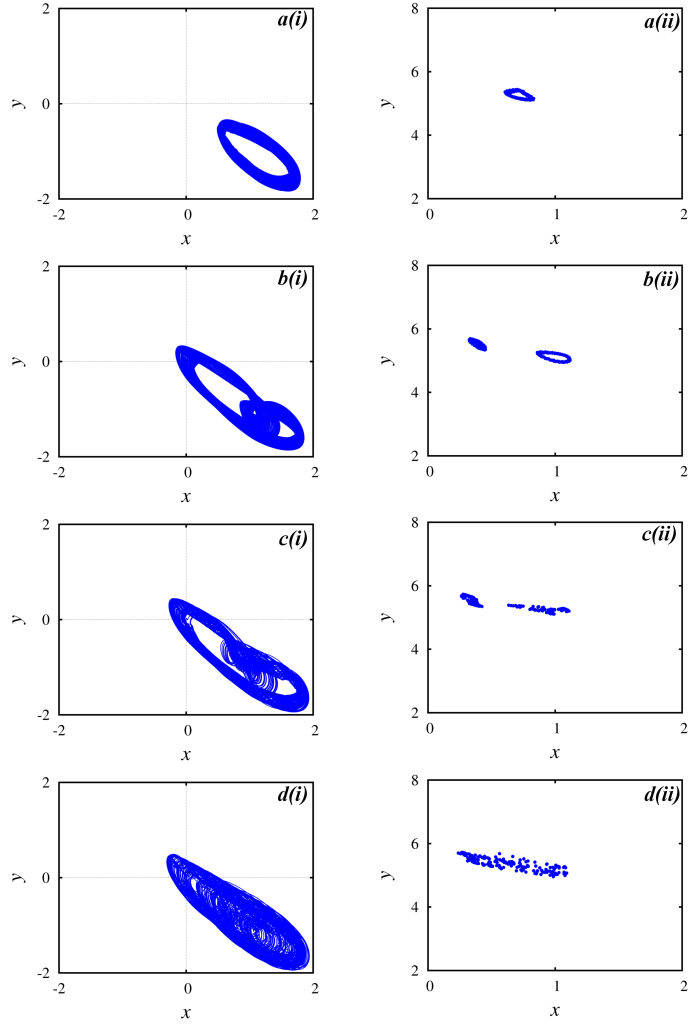}
}
\caption{Analytically obtained (a) phase portraits and their corresponding (b) poincare maps for (i) 1-torus at $f_2 = 0.2$, (ii) 2-torus at $f_2 = 0.225$, (iii) SNA at $f_2 = 0.2355$ and, (iv) Chaos at $f_2 = 0.236$.}
\label{fig:2}       
\end{figure}

With the circuit parameters taking the values $C=10~nF, L=18~mH, R_s=20~\Omega, R=1440~\Omega, \omega_1=23706.6~Hz$ and $\omega_2=7325.7~Hz$, the rescaled values of the normalized state variables takes the values, $\beta=1.152~, \nu=0.01388$, a=-1.0944, b=-0.5904, $\omega_1=2.1448,~ \omega_2=0.6627$ and $f_1 = 0.5184$. The amplitude of the second force $f_2$ has been taken as the control parameter. The SNA dynamics of the circuit exhibiting the {\emph{Heagy-Hammel}} and {\emph{fractalization}} routes has been studied analytically \cite{Sivaganesh2014}. The analytical solutions obtained for each piecewise linear region is summarized as follows.\\
Since the roots $m_{1,2}$ in the $D_{0}$ region are real and unequal, the fixed point $(0,0)$ corresponding to the $D_0$ region is a {\emph{saddle}} or {\emph{hyperbolic fixed point}}. The state variables $y(t)$ and $x(t)$ are
\begin{eqnarray}
y(t) =  && C_1 e^ {\alpha_1 t} + C_2 e^ {\alpha_2 t} + E_1 + E_2 \sin \omega_1 t + E_3 \cos \omega_1 t + E_4 \sin \omega_2 t + E_5 \cos \omega_2 t,
\label{eqn:6}
\end{eqnarray}
\begin{equation}
x(t) =  (\frac{1}{\beta})(- {\dot{y}} - \sigma y + f_1 sin( \omega_1 t) + f_2 sin( \omega_2 t)),
\label{eqn:7}
\end{equation}
The roots $m_{3,4}$ in the $ D_{\pm1}$ region are a pair of complex conjugates with a negative real part. Hence the fixed points $(\mp ((a-b) \sigma / \beta+b \sigma), \mp (\beta (b-a) / \beta+b \sigma))$ corresponding to the $D_{\pm1}$ regions is a { \emph{stable spiral fixed point }}. The state variables $y(t)$ and $x(t)$ are
\begin{eqnarray}
y(t) = && e^{u t}(C_3 \cos(vt) + C_4 \sin(vt)) + E_6 + E_7 \sin \omega_1 t   + E_8 \cos \omega_1 t + E_9 \sin \omega_2 t \nonumber \\
&& + E_0 \cos \omega_2 t\pm \Delta, 
\label{eqn:8}
\end{eqnarray}
\begin{equation}
x(t) =  (\frac{1}{\beta})(- {\dot{y}} - \sigma y + f_1 sin( \omega_1 t) + f_2 sin( \omega_2 t)),
\label{eqn:9}
\end{equation}
where $\Delta = \beta (b-a)$ and $+\Delta$, $-\Delta$ corresponds to $D_{+1}$ and $D_{-1}$ regions respectively.
The ratio of the frequencies, ($\frac {\omega_1}{\omega_2}$) has been found to be an integral multiple of the golden ratio, i.e.  $\frac {\omega_1}{\omega_2}= 2 (\frac{\sqrt{5}+1}{2})$. Hence, the system must exhibit SNA dynamics for a proper choice of the amplitudes of the external forcing terms $f_1$ and $f_2$. The SNA dynamics of the circuit exhibiting the HH route has been studied analytically through phase portraits and poincare maps. Fig.~\ref{fig:2} shows the phase portraits and their corresponding poincare maps for a 1-torus for $f_2 = 0.2$, a 2-torus for $f_2 = 0.225$ and a SNA followed by chaos in the range $0.15 < f_2 < 0.26$. Fig.~\ref{fig:2}(a(i))-\ref{fig:2}(a(iv)) represent the phase portraits of the Heagy-Hammel route in the $(x-y)$ phase plane. The correponding poincare maps of the HH route are shown in Fig. \ref{fig:2}(b). In the next section, we present the normalized state equations of the response system and explain the the analytical solutions obtained for the coupled system.

\section{Analytical Solution for Synchronization of SNAs}
\label{sec:2}

If we consider the system given by Eq.~(\ref{eqn:3}) acting as the drive system drives a response system with state variables $(x^{'}, y^{'}, \theta^{'}, \phi^{'}$) then the normalized state equations of the response system is given by
\begin{subequations}
\begin{eqnarray}
\dot x^{'}  &=&  y^{'} - g(x^{'}) + \epsilon (x-x^{'}), \\
\dot y^{'}  &=&  - \sigma y^{'} - \beta x^{'}+ f_3 sin(\theta^{'}) + f_4 sin(\phi^{'}), \\
\dot \theta^{'}  &=&  \omega_3, \\
\dot \phi^{'} &=& \omega_4,
\end{eqnarray}
\label{eqn:10}
\end{subequations}
where,
\begin{equation}
g(x^{'}) =
\begin{cases}
bx^{'}+(a-b) & \text{if $x^{'} \ge 1$}\\
ax^{'} & \text{if $|x^{'}|\le 1$}\\
bx^{'}-(a-b) & \text{if $x^{'} \le -1$}.
\end{cases}
\end{equation}
The circuit parameters of the response system takes the same value as that of the drive system. The drive and the response systems given by Eqs.~(\ref{eqn:3}) and (\ref{eqn:10}) could be used to obtain explicit analytical solutions for studying the synchronization dynamics.

\begin{figure}
\resizebox{0.6\textwidth}{!}{%
  \includegraphics{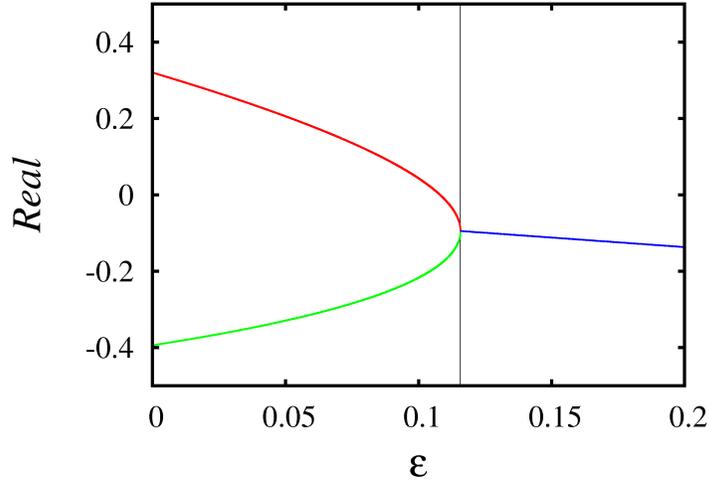}
}
\caption{Bifurcation of real eigenvalues as function of the coupling paramter in the $D^{*}_{0}$ region. The real eigenvalues transforms into complex conjugates for the coupling paramter $\epsilon >0.1157$.}
\label{fig:3}       
\end{figure}

From the state equations of the response system given by Eqs.~(\ref{eqn:10}), we observe that the dynamics of the response is influenced by the drive through the coupling parameter. Because the circuit equations are piecewise linear, each picewise linear region of the two sytems could be coupled together to get a new set of equations which could be solved for each region. The difference system obtained from Eqs.~(\ref{eqn:3}) and (\ref{eqn:10}) are
\begin{subequations}
\begin{eqnarray}
\dot {x^{*}}  = && y^{*} -  g(x^{*}) - \epsilon x^{*},\\
\dot {y^{*}}  = && -\sigma y^{*} - \beta x^{*} +f_1 sin(\omega_1 t) +f_2 sin(\omega_2 t) - f_3 sin(\omega_3 t) - f_4 sin(\omega_4 t), 
\end{eqnarray}
\label{eqn:12}
\end{subequations}
where $x^{*}$=$(x-x^{'})$, $y^{*}$=$(y-y^{'})$ and $g(x^{*}) = g(x) - g(x^{'})$ takes the values $a{x^{*}}$ or $b{x^{*}}$ depending upon the region of operation of the drive and response system. From the new set of state variables $x^{*}(t),~y^{*}(t)$, the state variables of the response system  $x^{'}(t),~y^{'}(t)$ could be written as
\begin{subequations}
\begin{eqnarray}
x^{'} &=& x -  x^{*}, \\
y^{'} &=& y -  y^{*}.
\label{eqn:13}
\end{eqnarray}
\label{eqn:13}
\end{subequations}
One can easily establish that a unique equilibrium point $ (x^{*}_0,y^{*}_0)$ exists for Eq.~\ref{eqn:12} in each of the following three subsets
\begin{equation}
\left.
\begin{aligned}
D^{*}_{+1} & =  \{ (x^{*},y^{*})| x^{*} \ge 1 \} P^{*}_+ = (0,0),\\
D^{*}_0 & =  \{ (x^{*},y^{*})|| x^{*} | \le 1 \}| O^{*} = (0,0),\\
D^{*}_{-1} & =  \{ (x^{*},y^{*})|x^{*} \le -1 \}| P^{*}_- = (0,0),\\
\end{aligned}
\right\}
\quad\text{}
\label{eqn:14}
\end{equation}
Form Eq.~(\ref{eqn:14}) it has been found that the origin $(0,0)$ is the fixed point for all the three piecewise linear regions of the difference system. In the first case, $g(x)$ and $g(x^{'})$ take the values  $a{x}$ and $a{x^{'}}$ respectively, which has been taken as the $D^{*}_0$ region of the difference system. The stability determining eigenvalues are calculated from the stability matrix
\begin{equation}
J^{*}_0 =
\begin{pmatrix}
-(a+\epsilon) &&& 1 \\
-\beta &&& -\sigma \\
\end{pmatrix},
\label{eqn:15}
\end{equation}
The eigenvalues in this region are real, negative and distinct for $\epsilon \le 0.1157$ while they are a pair of complex conjugates with negative real parts for  $\epsilon > 0.1157$. Hence, the stability of the fixed point transforms from a {\emph{stable node}} to a {\emph{stable spiral}} for  $\epsilon > 0.1157$.\\
In the second case, $g(x)$ and $g(x^{'})$ take the values  $bx \pm (a-b)$ and $bx^{'} \pm (a-b)$ respectively, which has been taken as the $D^{*}_{\pm1}$ regions of the difference system. The stability determining eigen values are calculated from the stability matrix
\begin{equation}
J^{*}_{\pm1} =
\begin{pmatrix}
-(b+\epsilon) &&& 1 \\
-\beta &&& -\sigma \\
\end{pmatrix},
\end{equation}
The eigenvalues in this region are found to be a pair of complex conjugates with a negative real part, for all the values of the coupling strength. Fig.~\ref{fig:3} shows the bifurcation of the real eigenvalues of the Jacobian matrix given by Eq.~(\ref{eqn:15}) in the $D^{*}_{0}$ as a function of the coupling parameter. The red and green lines show the two real roots while the blue line shows the real part of the complex conjugate roots. The tranformation of the eigenvalues at the critical value of the coupling paramter $\epsilon=0.1157$ indicates the mechanism of synchronization.


 When the coupling paramter $\epsilon=0$, the coupled systems become independent of each other. Hence the drive and the response systems given by Eqs.~(\ref{eqn:3}) and (\ref{eqn:10}) have the same solution for their state variables in all the three piecewise linear regions. The systems are operated with different set of initial conditions given by ($x_0=-0.5,~y_0=0.1$) and ($x^{'}_0 = 0.5,~y^{'}_0=0.11$). Owing to a difference in the initial conditions, the SNAs of the drive and response become unsynchronized for $\epsilon = 0$. The analytical solutions to the state variables in each of the three regions are as given in Eqs.~(\ref{eqn:6})-~(\ref{eqn:8}). \\

Now we present explicit analytical solutions for the dynamics of the response system for coupling strengths $\epsilon > 0$, leading to complete synchronization of the drive and the response. This is achieved by finding a solution to the normalized state variables of the difference system given by Eq.~(\ref{eqn:12}). The solution of those equations are, $ [x^{*} (t; t_0, x^{*}_0, y^{*}_0), ~y^{*}(t; t_0, x^{*}_0, y^{*}_0)]^T$ for which the initial conditions are written as $ (t, x^{*}, y^{*}) $ $ = (t_0, x^{*}_0, y^{*}_0) $. Since Eq.~(\ref{eqn:12}) is piecewise linear, the solution to each of the three piecewise linear regions can be obtained explicitly. 


In the $D^{*}_0$ region $g(x)$ and $g(x^{'})$ takes the values  $a{x}$ and $a{x^{'}}$ respectively. Hence the normalized equations obtained from Eqs.~(\ref{eqn:12}) are
\begin{subequations}
\begin{eqnarray}
\dot {x^{*}}  = && y^{*} - a x^{*} - \epsilon x^{*},\\
\dot {y^{*}}  = && -\sigma y^{*} - \beta x^{*} +f_1 sin(\omega_1 t) +f_2 sin(\omega_2 t) - f_3 sin(\omega_3 t) - f_4 sin(\omega_4 t), 
\end{eqnarray}
\label{eqn:17}
\end{subequations}
Differentiating Eq.~(\ref{eqn:17}b) with respect to time and using Eqs.~(\ref{eqn:17}a,~\ref{eqn:17}b) in the resultant equation, we obtain
\begin{eqnarray}
{\ddot y^{*}} + {A \dot y^{*}} + By^{*} = && (a+\epsilon) f_1~sin \omega_1 t +  (a+\epsilon) f_2~sin \omega_2 t  - (a+\epsilon) f_3~sin \omega_3 t - (a+\epsilon) f_4~sin \omega_4 t \nonumber \\ 
&& + f_1 \omega_1~cos \omega_1 t  +  f_2 \omega_2~cos \omega_2 t - f_3 \omega_3~cos \omega_3 t  -  f_4 \omega_4~cos \omega_4 t,
\label{eqn:18}
\end{eqnarray}
where, $ A = \sigma + a + \epsilon$ and $ B = \sigma(a+\epsilon)+\beta$. 
The roots of Eq.~(\ref{eqn:18}) are ${m_{1,2}} =  \frac{-(A) \pm \sqrt{(A^{2}-4B)}} {2}$.
From stability analysis, for $\epsilon \le 0.1157$, $(A^{2} > 4B)$, and hence the roots $m_1$ and $m_2$ are real, negative and distinct. Using the method of undetermined coefficients, the general solution to Eq.~(\ref{eqn:18}) can be written as
\begin{eqnarray}
y^{*}(t) = && C_1 e^ {m_1 t} + C_2 e^ {m_2 t} + E_1 sin(\omega_1 t) + E_2 cos(\omega_1 t) + E_3 sin(\omega_2 t) + E_4 cos(\omega_2 t) +  E_5 sin(\omega_3 t) \nonumber \\
&& + E_6 cos(\omega_3 t) + E_7 sin(\omega_4 t) + E_8 cos(\omega_4 t),
\label{eqn:19}
\end{eqnarray}
where $C_1$ and $C_2$ are integration constants and
\begin{subequations}
\begin{eqnarray}
E_1  =&&  \frac {f_1  {\omega_1} ^2 (A-a-\epsilon) + f_1 B(a+\epsilon)}{A^2 {\omega_1} ^2 + (B-{\omega_1} ^2)^2},   \\
E_2  =&& \frac {f_1  \omega_1 ((B-{\omega_1} ^2)-A(a+\epsilon ))}{A^2 {\omega_1} ^2 + (B-{\omega_1} ^2)^2},   \\
E_3  =&&  \frac {f_2  {\omega_2} ^2 (A-a-\epsilon) + f_2 B(a+\epsilon)}{A^2 {\omega_2} ^2 + (B-{\omega_2} ^2)^2},   \\
E_4  =& & \frac {f_2  \omega_2 ((B-{\omega_2} ^2)-A(a+\epsilon ))}{A^2 {\omega_2} ^2 + (B-{\omega_2} ^2)^2},  \\
E_5  =&&  \frac {f_3  {\omega_3} ^2 (a+\epsilon-A) - f_3 B(a+\epsilon)}{A^2 {\omega_3} ^2 + (B-{\omega_3} ^2)^2},  \\
E_6  =&& \frac {f_3  \omega_3 (A(a+\epsilon )-(B-{\omega_3} ^2))}{A^2 {\omega_3} ^2 + (B-{\omega_3} ^2)^2},  \\
E_7  =&&  \frac {f_4  {\omega_4} ^2 (a+\epsilon-A) - f_4 B(a+\epsilon)}{A^2 {\omega_4} ^2 + (B-{\omega_4} ^2)^2},  \\
E_8  =&& \frac {f_4  \omega_4 (A(a+\epsilon )-(B-{\omega_4} ^2))}{A^2 {\omega_4} ^2 + (B-{\omega_4} ^2)^2},
\end{eqnarray}
\label{eqn:20}
\end{subequations}

The constants $C_1$ and $C_2$ are given as, 
\begin{subequations}
\begin{eqnarray}
C_1 =  &&\frac{e^ {- m_1 t_0}} {m_1 - m_2} \{ (-\sigma{ y^{*}_0}-\beta{ x^{*}_0}-m_2{ y^{*}_0}) - ( E_1 \omega_1 - m_2 E_2) cos \omega_1 t_0 + (E_2 \omega_1 + m_2 E_1+ f_1) sin \omega_1 t_0  \nonumber \\
&&  -(E_3 \omega_2 - m_2 E_4) cos \omega_2 t_0 + (E_4 \omega_2 + m_2 E_3 + f_2) sin \omega_2 t_0 - ( E_5 \omega_3 - m_2 E_6) cos \omega_3 t_0 \nonumber \\
&& + (E_6 \omega_3 + m_2 E_5- f_3) sin \omega_3 t_0 - (E_7 \omega_4 - m_2 E_8) cos \omega_4 t_0 + (E_8 \omega_4 + m_2 E_7 - f_4) sin \omega_4 t_0 \} \nonumber \\
\\
C_2 =  &&\frac{e^ {- m_2 t_0}} {m_2 - m_1} \{ (-\sigma{ y^{*}_0}-\beta{ x^{*}_0}-m_1{ y^{*}_0}) - ( E_1 \omega_1 - m_1 E_2) cos \omega_1 t_0 + (E_2 \omega_1 + m_1 E_1+ f_1) sin \omega_1 t_0  \nonumber \\
&&  -(E_3 \omega_2 - m_1 E_4) cos \omega_2 t_0 + (E_4 \omega_2 + m_1 E_3 + f_2) sin \omega_2 t_0  - ( E_5 \omega_3 - m_1 E_6) cos \omega_3 t_0 \nonumber \\
&& + (E_6 \omega_3 + m_1 E_5- f_3) sin \omega_3 t_0 - (E_7 \omega_4 - m_1 E_8) cos \omega_4 t_0 + (E_8 \omega_4 + m_1 E_7 - f_4) sin \omega_4 t_0 \} \nonumber \\
\end{eqnarray}
\label{eqn:21}
\end{subequations}
Differentiating Eq.~(\ref{eqn:18}) and using it in Eq.~(\ref{eqn:17}b) we get
\begin{eqnarray}
x^{*}(t) = && \frac{1}{\beta}(\dot{y^{*}} - \sigma y^{*} + f_1 sin(\omega_1 t) +f_2 sin(\omega_2 t) - f_3 sin(\omega_3 t) - f_4 sin(\omega_4 t)). 
\label{eqn:22}
\end{eqnarray}
From the results of $y^{*}(t),~x^{*}(t)$ obtained from Eqs.~(\ref{eqn:19}),~(\ref{eqn:22}) and $y(t),~x(t)$ obtained from Eqs.~(\ref{eqn:6}),~(\ref{eqn:7}), $x^{'}(t)$ and $y^{'}(t)$ can be obtained from Eqs.~(\ref{eqn:13}).

For $\epsilon > 0.1157$, $(A^{2} < 4B)$ and the roots $m_1$ and $m_2$ are a pair of complex conjugates with a negative real part given as $m_{1,2} = u \pm iv$, with $u=\frac{-A}{2}$ and $v=\frac{\sqrt(4B-A^{2})}{2}$.
The general solution to eq.(13) can be written as,
\begin{eqnarray}
y^{*}(t) =  && e^ {ut} (C_1 cosvt+ C_2 sinvt)+ E_1 sin(\omega_1 t) + E_2 cos(\omega_1 t) + E_3 sin(\omega_2 t) + E_4 cos(\omega_2 t) \nonumber \\
&& +   E_5 sin(\omega_3 t) + E_6 cos(\omega_3 t) + E_7 sin(\omega_4 t) + E_8 cos(\omega_4 t),
\label{eqn:23}
\end{eqnarray}
where the constants $E_1,E_2,E_3,E_4,E_5,E_6,E_7,E_8$ are as given in given in Eq.~(\ref{eqn:20}).
Differentiating Eq.~(\ref{eqn:23}) and using it in Eq.~(\ref{eqn:17}b) we get,
\begin{eqnarray}
x^{*}(t) = && (\frac{1}{\beta})(-\dot{y^{*}} - \sigma y^{*}+ f_1 sin(\omega_1 t) +f_2 sin(\omega_2 t) - f_3 sin(\omega_3 t) - f_4 sin(\omega_4 t)). 
\label{eqn:24}
\end{eqnarray}
The constants $C_1$ and $C_2$ are given as,
\begin{subequations}
\begin{eqnarray}
C_1 =  && \frac{e^ {- u t_0}} {v} \{((\sigma { y^{*}_0} +\beta { x^{*}_0}+u { y^{*}_0})sinvt_0 + v { y^{*}_0} cosvt_0)  +((E_1 \omega_1 - u E_2) sinvt_0 - vE_2 cos vt_0)cos \omega_1 t_0 \nonumber \\
&&- ((E_2 \omega_1+u E_1+f_1) sinvt_0+v E_1 cosvt_0) sin \omega_1 t_0  +((E_3 \omega_2 - u E_4) sinvt_0 - vE_4 cos vt_0)cos \omega_2 t_0  \nonumber \\
&&- ((E_4 \omega_2+u E_3+f_2) sinvt_0+v E_3 cosvt_0) sin \omega_2 t_0  +((E_5 \omega_3 - u E_6) sinvt_0 - vE_6 cos vt_0)cos \omega_3 t_0  \nonumber \\
&&- ((E_6 \omega_3+u E_5-f_3) sinvt_0+v E_5 cosvt_0) sin \omega_3 t_0  +((E_7 \omega_4 - u E_8) sinvt_0 - vE_8 cos vt_0)cos \omega_4 t_0 \nonumber \\
&& - ((E_8 \omega_4+u E_7-f_4) sinvt_0+v E_7 cosvt_0) sin \omega_4 t_0,
\\
C_2 =  && \frac{e^ {- u t_0}} {v} \{((-\sigma { y^{*}_0} -\beta { x^{*}_0}-u { y^{*}_0})cos vt_0 + v { y^{*}_0} sin t_0) -((E_1 \omega_1 - u E_2) cos vt_0 + vE_2 sin vt_0)cos \omega_1 t_0  \nonumber \\
&&+ ((E_2 \omega_1+u E_1+f_1) cos vt_0-v E_1 sin vt_0) sin \omega_1 t_0  -((E_3 \omega_2 - u E_4) cos vt_0 + vE_4 sin vt_0)cos \omega_2 t_0  \nonumber \\
&&+ ((E_4 \omega_2+u E_3+f_2) cos vt_0-v E_3 sin vt_0) sin \omega_2 t_0 -((E_5 \omega_3 - u E_6) cos vt_0 + vE_6 sin vt_0)cos \omega_3 t_0  \nonumber \\
&&+ ((E_6 \omega_3+u E_5-f_3) cos vt_0-v E_5 sin vt_0) sin \omega_3 t_0 -((E_7 \omega_4 - u E_8) cos vt_0 + vE_8 sin vt_0)cos \omega_4 t_0  \nonumber \\
&&+ ((E_8 \omega_4+u E_7-f_4) cos vt_0-v E_7 sin vt_0) sin \omega_4 t_0  
\end{eqnarray}
\label{eqn:25}
\end{subequations}
From the results of $y^{*}(t),~x^{*}(t)$ obtained from Eqs.~(\ref{eqn:23}),~(\ref{eqn:24}) and $y(t),~x(t)$ obtained from Eqs.~(\ref{eqn:6}),~(\ref{eqn:7}), $x^{'}(t)$ and $y^{'}(t)$ can be obtained from Eqs.~(\ref{eqn:13}).

In the $D^{*}_{\pm1}$ region $g(x)$ and $g(x^{'})$ takes the values  $b{x}\pm (a-b)$. Hence the normalized equations obtained from Eqs.~(\ref{eqn:12}) are
\begin{subequations}
\begin{eqnarray}
\dot {x^{*}} & = & y^{*} - (b+\epsilon){x^{*}} \\
\dot {y^{*}}  = && -\sigma y^{*} - \beta x^{*} +f_1 sin(\omega_1 t) +f_2 sin(\omega_2 t) - f_3 sin(\omega_3 t) - f_4 sin(\omega_4 t), 
\end{eqnarray}
\label{eqn:26}
\end{subequations}
Differentiating Eq.~(\ref{eqn:26}b) with respect to time and using Eqs.~(\ref{eqn:26}a, \ref{eqn:26}b) in the resultant equation, we obtain
\begin{eqnarray}
{\ddot y^{*}} + {C \dot y^{*}} + D y^{*} = && (b+\epsilon) f_1~sin \omega_1 t +  (b+\epsilon) f_2~sin \omega_2 t - (b+\epsilon) f_3~sin \omega_3 t - (b+\epsilon) f_4~sin \omega_4 t \nonumber \\ 
&& + f_1 \omega_1~cos \omega_1 t  +  f_2 \omega_2~cos \omega_2 t - f_3 \omega_3~cos \omega_3 t  -  f_4 \omega_4~cos \omega_4 t,
\label{eqn:27}
\end{eqnarray}
where, $ C = \sigma + b + \epsilon$ and $ D = \sigma(b+\epsilon)+\beta$. 
From the stability analysis we could confirm that the roots $m_3$ and $m_4$ are a pair of complex conjugates given as $m_{3,4} = u \pm iv$, with $u=\frac{-C}{2}$ and $v=\frac{\sqrt(4D-C^{2})}{2}$, for all values of the coupling strength. Hence the state variables, $(y^{*}(t),x^{*}(t))$ can be written as 
\begin{eqnarray}
y^{*}(t) = && C_1 e^ {m_1 t} + C_2 e^ {m_2 t} + E_1 sin(\omega_1 t) + E_2 cos(\omega_1 t) + E_3 sin(\omega_2 t) + E_4 cos(\omega_2 t) +  + E_5 sin(\omega_3 t) \nonumber \\
&& + E_6 cos(\omega_3 t) + E_7 sin(\omega_4 t) + E_8 cos(\omega_4 t),
\label{eqn:28}
\end{eqnarray}
\begin{eqnarray}
x^{*}(t) = && (\frac{1}{\beta})(-\dot{y^{*}} - \sigma y^{*}+ f_1 sin(\omega_1 t) +f_2 sin(\omega_2 t) - f_3 sin(\omega_3 t) - f_4 sin(\omega_4 t)). 
\label{eqn:29}
\end{eqnarray}
The constants $E_1, E_2, E_3, E_4,E_5,E_6, E_7, E_8$ and $C_3, C_4$ are the same  as the contants given in Eq.~(\ref{eqn:20}) and~(\ref{eqn:25}) respectively, except that the constants $A,~B$ are replaced with $C,~D$. 
From the results of $y^{*}(t),~x^{*}(t)$ obtained from  Eqs.~(\ref{eqn:28}),~(\ref{eqn:29}) and $y(t),~x(t)$ obtained from  Eqs.~(\ref{eqn:8}),~(\ref{eqn:9}), $x^{'}(t)$ and $y^{'}(t)$ can be obtained from Eq.~(\ref{eqn:13}).
Now let us briefly explain how the solution can be generated in the $(x^{'}-y^{'})$ phase space. With the {\emph{time (t)}} being considered as the independent variable, the state variables evolves within each piecewise linear region depending upon its initial values. If we start with the initial conditions $x^{*}(t=0) = x^{*}_0,~y^{*}(t=0) = y^{*}_0$ in the $D^{*}_0$ region at time $t=0$, the arbitrary constants $C_1$ and $C_2$ get fixed. Thus $x^{*}(t)$ evolves as given by Eq.~(\ref{eqn:22}) up to either $t=T_1$, when $x^{*}(T_1)=1$ or $t = T^{'}_1$ when $x^{*}(T^{'}_1) = -1$. The next region of operation $(D^{*}_{+1}$ or $D^{*}_{-1})$ thus depends upon the value of $x^{*}$ in the $D^{*}_{0}$ region at that instant of time. As the trajectory enters into the next region of interest, the arbitrary constant corresponding to that region could be evaluated, with the initial conditions to that region being either ($x^{*}_{0}(T_1),~y^{*}_{0}(T_1)$) or  ($x^{*}_{0}(T^{'}_1),~y^{*}_{0}(T^{'}_1)$). During each region of operation, the state variables of the response system evolves as $x^{'}(t) = x(t) - x^{*}(t)$ and $y^{'}(t) = y(t) - y^{*}(t)$, respectively. The procedure can be continued for each successive crossing. In this way, the explicit solutions can be obtained in each of the regions $D^{'}_0$, $D^{'}_{\pm1}$ of the response system. The solution obtained in each region has been matched across the boundaries and used to generate the dynamics of the response system. \\

\begin{figure}
\resizebox{0.6\textwidth}{!}{%
  \includegraphics{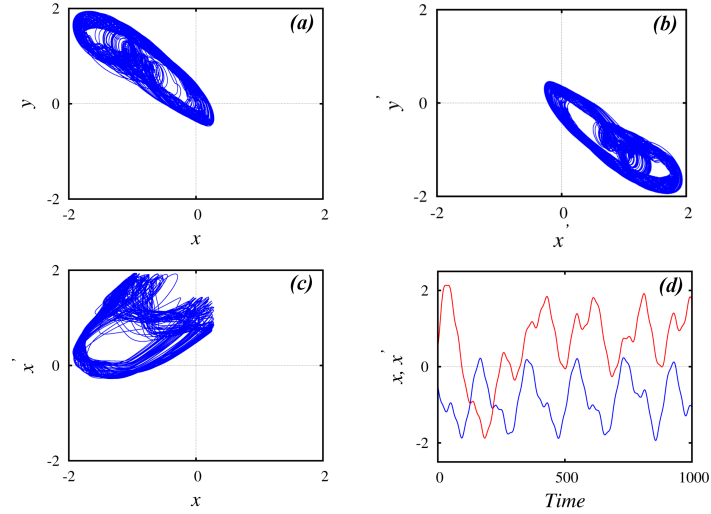}
}
\caption{Unsynchronized state of coupled SNAs for $\epsilon = 0$. (a) SNA of drive in $(x-y)$ phase space, (b) SNA of response in $(x^{'}-y^{'})$ phase space, (iii) Unsynchronized motion in $(x-x^{'})$ phase space and (iv) Time series of the state variables $x$ (blue line) and $x^{'}$ (red line).}
\label{fig:4}       
\end{figure}

\begin{figure}
\resizebox{0.6\textwidth}{!}{%
  \includegraphics{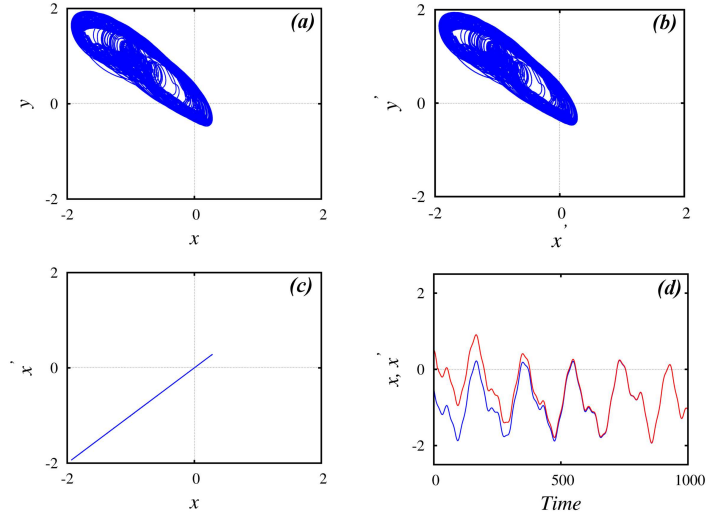}
}
\caption{Synchronized motion of coupled SNAs for $\epsilon = 0.1158$. (a) SNA of drive in $(x-y)$ phase space, (b) SNA of response in $(x^{'}-y^{'})$ phase space, (iii) Synchronized state in $(x-x^{'})$ phase space and (iv) Time series of the state variables $x$ (blue line) and
$x^{'}$ (red line).}
\label{fig:5}       
\end{figure}

\begin{figure}
\resizebox{0.6\textwidth}{!}{%
  \includegraphics{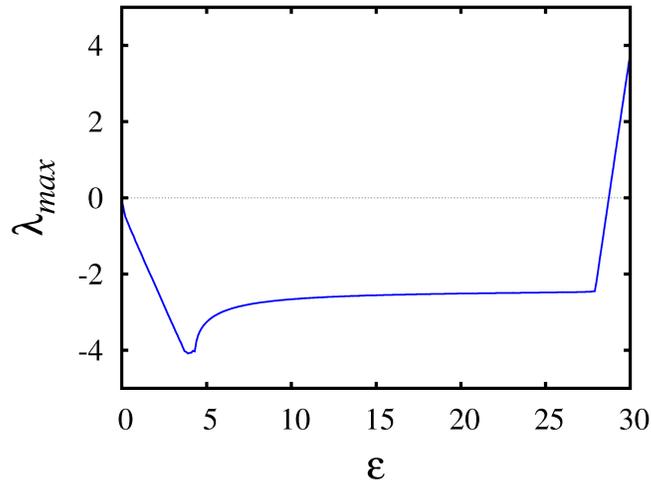}
}
\caption{MSF ($\lambda_{max}$) as a function of the coupling parameter ($\epsilon$) for the x-coupled system. The broad negative value region indicates the stable synchronized state of the coupled system.}
\label{fig:6}       
\end{figure}

The analytical solutions obtained above for the response system can be used to explain the phenomena of complete synchronization through phase portraits and timeseries analysis. The initial conditions of the drive and the response systems are so chosen such that the SNAs of the two system exist in different regions of phase space. The initial conditions of the drive and response systems has been fixed as ($x_0$=-0.5, $y_0$=0.1) and ($x^{'}_0$ = 0.5, $y^{'}_0$=0.11) respectively. Fig.~\ref{fig:4}(a),~\ref{fig:4}(b) shows the phase portraits of the SNAs of drive and response systems. The unsynchronized state of the coupled system for the coupling parameter $\epsilon=0$, in the $(x-x{'})$ phase plane and their corresponding trajectory in the $x^{*}=x-x^{'}$ plane is shown in Fig.~\ref{fig:4}(c) and \ref{fig:4}(d) respectively. For $\epsilon > 0.1157$, the fixed point of the $D^{*}_{0}$  region transform into a {\emph{stable spiral}} indicating the asymptotic convergence of trajectories towards the origin, within the synchronization manifold. The phase portraits of the drive and the response systems for the value of the coupling parameter $\epsilon = 0.1158$ is shown in Fig.~\ref{fig:5}(a),~\ref{fig:5}(b), respectively. Fig.~\ref{fig:5}(c) shows the complete synchronization of the coupled systems in the $(x-x{'})$ phase plane and their corresponding trajectory in the $x^{*}=x-x^{'}$ plane as in Fig.~\ref{fig:5}(d). From the phase portraits and the time series plots obtained it could be inferred that for the coupling parameter taking the value $\epsilon=0.1158$, the response system which is operating with a different set of initial condition and for a different value of external periodic force, completely synchronizes with the drive. The stability of the synchronized state for the x-coupled system is confirmed through the {\emph{MSF}} obtained from the normalized state equations \ref{eqn:3} and \ref{eqn:10}. The MSf for the coupled SNA system is as shown in Fig.~\ref{fig:6}. The figure shows a broader stable synchronized region within $0.1157<\epsilon<28.5$ indicated by the negative values of $\lambda_{max}$, confirming the synchronization of the coupled system.

\section{Conclusion}

We have presented in this paper an explicit analytical solution to the normalized state equations of coupled quasiperiodically-forced systems exhibiting SNAs in their dynamics. The solutions thus obtained have been used to explain the phenomenon of complete synchronization observed in the coupled system through phase portraits. Analytical solutions for the synchronization of simple chaotic systems have been reported recently \cite{Sivaganesh2015,Venkatesh2016}. To the best of our knowledge, this is the first time that synchronization of SNAs is studied analytically.

\pagebreak

\end{document}